\documentclass[10pt]{article}

\usepackage{amsmath}
\usepackage{amssymb}
\usepackage{graphicx}
\usepackage{epstopdf}
\usepackage[dvips]{epsfig}
\usepackage{url}
\usepackage{verbatim}
\usepackage{natbib}
\usepackage[left=7mm,right=15mm]{geometry} 
\usepackage{vmargin} 
\usepackage{tabularx}
\usepackage{hyperref}
\usepackage{enumerate}

\title{\Large{Probabilistic Mortality Forecasting with Varying Age-Specific Survival Improvements}}
\author{\large{Christina Bohk and Roland Rau}\\ \large{University of Rostock}}
\date{}  

\begin{document}

\maketitle

\section*{Summary}

We propose a probabilistic mortality forecasting model that can be applied to derive forecasts for populations with regular and irregular mortality developments.
Our model (1) uses rates of mortality improvement to model dynamic age patterns of mortality change and it can
(2) optionally complement the mortality trend of a country of interest with that of at least one reference country.
Retrospective mortality forecasts for British and Danish women from 1991 to 2011 suggest that our model can generate smaller forecast errors than other widely accepted approaches like, for instance, the Lee-Carter model or the UN Bayesian approach. 

%\keywords{Bayesian methods, Demography, Life expectancy at birth, Markov chain Monte Carlo, Mortality forecasting, Rates of mortality improvement}

\section{Introduction}

Few activities of demographers receive as much attention from society
as population forecasts. The recent track record is not praiseworthy,
though. Despite longer time series, presumably better 
methodology and more computing power, \cite{keilmanPDR}
concluded for European countries that ``Demographic Forecasts Have Not
Become More Accurate Over the Past 25 Years''. Even projecting only
mortality --- as one of the three parameters besides fertility and
migration that determine the size of a population and shape its
age-structure --- does not yield satisfactory results
\citep[p.~146]{keilmanPDR}. 

The canonical Lee-Carter method \citep{Carter1992, leemiller,
leecarter, Lee1992} generates robust mortality forecasts with a
relatively simple and parsimonous principal component model that
captures age and period effects. 
Numerous refinements and extensions of the original Lee-Carter model
were proposed to increase forecast accuracy (see for an overview
\cite{Booth2006,Booth2008,Shang2011,Butt2010,Shang2012}); 
for instance, Renshaw and Haberman \citeyearpar{Renshaw2003, Renshaw2006}
generalized the Lee-Carter model and included additional terms to
capture period and/or cohort effects optionally. 
Although these refinements increased forecasting accuracy of the original
Lee-Carter model significantly, 
it is still a challenging task to forecast unsteady mortality
developments. This is also true for more flexible approaches such as
the $P$-spline approach of Currie et al.~\citeyearpar[p.~297]{curriemortality} who state that they failed ``to
predict accurately the fall in mortality rates''.

The lack of adequate methods to predict mortality stands in stark
contrast to its importance. Preston and Stokes \citeyearpar[p.~227]{prestonstokes2012}
showed that population aging in more developed countries is primarily
the outcome of improved chances of survival. Few areas of public
policy remain unaffected by the increasing number of the elderly in a
population as a result of decreasing mortality. Financing old-age
pensions and health care as well as the provision of long-term care
are only the tip of the iceberg.
However, not only public policy is affected. Private 
companies also require reliable estimates for future mortality. Pension
funds represent the most obvious example \citep{OECD2011, Soneji2012}. 

Our contribution is a novel combination of modern approaches to forecast mortality:
\begin{enumerate}[(a)]
\item \emph{Rates of mortality improvement as the core forecasting component}
Mortality is typically forecasted by extrapolating death rates. 
For instance, the popular Lee-Carter model \citep{Carter1992,
  leemiller,leecarter, Lee1992} as well as many of its extensions
extrapolate past trends of age-specific mortality on the log scale. 
As a methodological advancement, very recent studies
\citep{Mitchell2013, Haberman2012} forecast the rates of mortality
improvement instead of the death rates.  For instance, \cite{Mitchell2013} and \cite{Haberman2012}
take the original (and extended) Lee-Carter framework,  but they
replace the log death rates with their corresponding rates of
improvement to  mortality.  Although these
approaches apply similar predictor structures, they differ in their
definition of the change of mortality: \cite{Mitchell2013} use period incremental mortality improvements,
whereas \cite{Haberman2012} use scaled mortality
improvement rates, ranging between -2 and 2. Both approaches
\citep{Mitchell2013,Haberman2012} argue that they yield better forecasting results than
the original Lee-Carter model and some of its variants. We also use
the rates of improvement rather 
than the level of mortality, defining the rates of mortality
improvement as the time-derivative of age-specific death rates. What
is the demographic rationale to use the first derivative of death
rates over time instead of the death rates themselves? In our opinion,
there are two main reasons: First, while the actual level of mortality
determines current life expectancy, it is the age-specific rate of change
that determines the development into the future. It has been
demonstrated in the past that this pace of survival improvements is
rather independent of the current level of mortality
\citep{vaupelremark, Kannisto1994}. Secondly, it is now well
established that life expectancy is rising for more than 170 years
(e.g. \citep{Oeppen2002, White2002, Tuljapurkar2000,
  Vallin2009}). Despite the linear pattern in the increase of life
expectancy, mortality did not decrease at all ages
simultaneously. Infant and childhood ages contributed most to the
increase in life expectancy in the 19th and the early 20th century,
whereas nowadays life expectancy rises primarily because of
reductions in mortality among the oldest-old
\citep{Christensen2009}. Death rates are falling now at very advanced
ages (e.g.~above~85 and even above 90), where mortality was often
considered to be fixed \citep{pdrarticle, jwvnature}, at such a rapid
pace that even purely data-driven models were unable to capture the
trend (e.g.~\cite{curriemortality}). We expect that modeling ``rates
of mortality improvement'' instead of death rates allows us to 
capture this age-shift to project mortality trends more accurately.

\item \emph{Complement mortality trends with expert opinion} Recent studies \citep{Li2005,Cairns2011} give evidence that it can be 
advantageous to forecast mortality of multiple countries jointly. For
instance, \cite{Cairns2011} proposed a Bayesian model to
generate consistent mortality forecasts for two populations, whereas
\cite{Li2005} extended the original Lee-Carter model to
forecast mortality of a single country with a shared trend among a
group of countries. In our proposed model, we take the mortality trend
of one or more reference countries into account, too, to complement
the mortality trend of a single country of interest in the long
run. This is especially useful when, for instance, the increase of
life expectancy at birth in a country of interest stagnates in the
base period, but is expected to continue or accelerate in the forecast
years. A purely data-driven extrapolation of such a sluggish mortality
trend could underestimate the actual mortality decline in the forecast
years and could, therefore, induce large forecast errors. To address
this issue, we think it is worthwhile to supplement the extrapolated
trend with a mortality schedule of at least one other reference
country. The selection of appropriate reference countries is thereby
due to expert judgment. The rationale behind adopting mortality trends
of reference populations is, according to \cite{Li2005} as
well as \cite{Cairns2011}, to avoid implausibly
diverging trends in mortality for a cluster of comparable countries.

\item \emph{Probabilistic approach} We propose to forecast age-specific
  mortality with a Bayesian model. Such a model can combine different
  sources of information like past trends, theories, and expert
  judgment regarding the future development of the rates of mortality
  improvement. Furthermore, our Bayesian mortality forecasts incorporate uncertainty inherently,
  i.\,e.~our model automatically captures and quantifies forecast uncertainty with probability statements.
  This is particularly important because
``[t]he demographic future of any country
  is uncertain. There is not just one possible future, but many. Some
  of these are more probable than others''
  \cite[p.~410]{keilmann2002}. Forecasting models have traditionally
  been deterministic, i.\,e.~they represent a ``what-if''
  scenario. Such an approach does not allow to assess how certain a 
  given scenario is (or multiple scenarios are). Confidence intervals,
  as provided by probabilistic models, though, give the user of a
  forecast some insights about the most likely boundaries of an estimate.
  Nico Keilman \citeyearpar{keilmannature, keilmann2002}
  differentiates between three methods of probabilistic forecasting:
  time-series extrapolation, expert judgment and extrapolation of
  historical forecast errors. Our model is a combination of the first
  two: We extrapolate past rates of mortality improvement, combined
  with constraints on the development of mortality derived from
  expert judgment. 

  Despite the early contribution of T\"{o}rnqvist in the middle of the 20th century \citep{AlhoSpencer2005, Tornqvist1948},
  generating demographic forecasts with probability statements and with Bayesian methods is still relatively new.
  For instance, \cite{Girosi2008} use a Bayesian model to forecast age-specific mortality (by cause of death) as a dependent variable
  that is influenced by multiple explanatory variables like sex, location, and GDP.  
  \cite{Raftery2013}, \cite{Chunn2010} and \cite{Alkema2011} revise the traditional deterministic approaches from the UN to forecast
  life expectancy at birth and the total fertility rate.
  They integrate the well-established double logistic function in a Bayesian model that combines country-specific and overall country information with a time series approach.  
  \cite{Czado2005}, \cite{Pedroza2006} as well as \cite{Kogure2010} transform the Lee-Carter model to a Bayesian approach
  to address problems with, for instance, erratic data, projection uncertainty, and missing data.
  \cite{Billari2012, Billari2013} propose a Supra-Bayesian model to forecast a population probabilistically,
  using opinions from experts regarding the future development of vital rates as input data in order to determine, e.\,g., their (expected) median values, marginal variability and correlation structure across time.
  \cite{Bijak2010} combine quantitative and qualitative information in a Bayesian model to forecast immigration for selectd European countries over a short projection horizon. 
  \cite{Abel2010} apply Bayesian time series models to forecast total population size.
However, none of those Bayesian approaches that forecast mortality can
sufficiently capture the flexible age schedule of mortality change so
far, and may induce, therefore, forecast errors.
\end{enumerate}

The overall aim of our paper is to propose a model that can generate
more accurate mortality forecasts, even in the presence of challenging
forecasting conditions, than other methods. 
Those methods often fail to capture unsteady mortality developments.
For instance, life expectancy during the 1980s and early 1990s did not
increase for women in Denmark as it did in many other
countries; instead, it almost stagnated.
A simple extrapolation of this development would underestimate the
progress in mortality that has actually been observed thereafter. 
To circumvent such methodological problems, we  use (1) flexible age
schedules of mortality change with the rates of mortality improvement 
in conjunction with (2) the mortality trend of at least one other
reference country in order to complement the purely extrapolated
mortality trend of the country of interest.  
In an application, we show that these two methodological features
actually allow us to forecast the mortality of Danish females more
accurately than other widely accepted models like the canonical Lee-Carter model or the UN Bayesian approach.\\

The remainder of our paper is organized as follows:  After formally
introducing our proposed mortality forecasting model, i.\,e.~its
input, core and output, in the next section, we apply it to forecast
mortality for British and Danish women thereafter---two populations
with divergent mortality developments in the recent past:  British
women experienced a regular mortality development with a rather stable
increase in life expectancy at birth,  while Danish women experienced
an irregular mortality development with a rather unstable increase in
life expectancy at birth.  To test and evaluate the performance of our
model under such diverse forecasting conditions, we compare its output
with that of the Lee-Carter model \citep{Carter1992}, with some
enhancements of the Lee-Carter model
\citep{Renshaw2003,Renshaw2006,Li2005}, with the $P$-spline appoach
\citep{curriemortality}, with the UN Bayesian approach \citep{Raftery2013} as well as with the Eurostat \emph{EUROPOP2010} forecast \citep{EC2011}.
Finally, we summarize and discuss our main findings in the last section.           

\section{Methodology}
\label{sec:forecastingModel}

Life expectancy at birth increases in many countries, but not necessarily at the same pace.
Irrespective of the level of mortality, periods of strong survival improvements can be followed by periods of weaker improvements and even by periods of stagnation or mortality deterioration.
Our goal is to propose a model,
which can forecast regular as well as irregular mortality developments.
To capture such diverse mortality developments for each country with only one model,
it contains several parameters to adjust to different forecasting conditions:
First, we use rates of mortality improvement (instead of death rates) to catch potential dynamic age shifts,
which are induced by increasing survival improvements for older ages in many highly developed countries.
Second, we optionally complement the mortality trend in a country of interest with those of selected reference countries;
this option enables our model to include changes in extrapolated long-time trends (for the country of interest) that are expected to occur in the forecast years.
Third, we implement our model in a Bayesian framework to capture inherent uncertainty of mortality forecasts.\\

\begin{figure}[!ht]
\centering
\makebox{\includegraphics[width=1\textwidth]{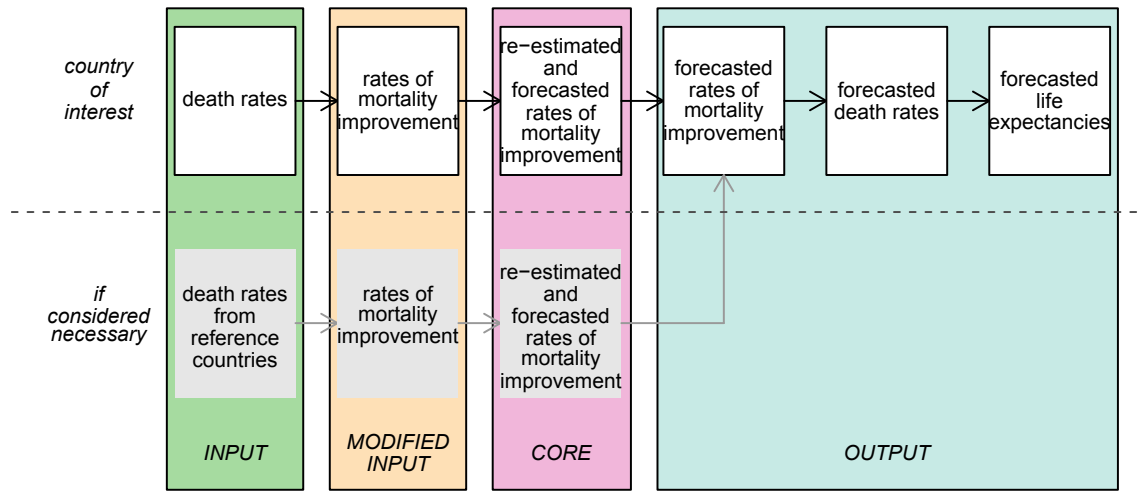}}
\caption{\label{fig:MethodPlot}Structure of our proposed mortality forecasting model:
Based on death rates as input, our model estimates rates of mortality
improvement, which are the core component of our approach, to forecast
death rates and, ultimately, life expectancies. If the mortality trend
in the country of interest appears implausible to continue, we can
optionally complement it with the mortality trend of at least one reference country.}
\end{figure}

Our model consists of three main parts, namely the model input, the
core and the model output (see Fig.~\ref{fig:MethodPlot}). 
We describe each model component in detail in the following subsections. \\  

\subsection{Model Input}

Our model requires death rates by age and calendar time  as input.
Such detailed mortality data are available for many countries in the Human Mortality Database \citeyearpar{hmd2013}.
To gain information about the age pattern of mortality change with time, we calculate the rates of mortality improvement for each age-specific death rate over time $\rho\left(x,t\right)$:

\begin{equation}\label{estiromi}
\rho\left(x,y\right) = - \log \left(\frac{m\left(x,y\right)}{m\left(x,y-1\right)}\right),
\end{equation}

where $m(x,y)$ denotes the death rate at age $x$ in year $y$.
Equation \ref{estiromi} is a transformation of the standard equation to estimate the rate of growth ($x_t=x_0 e^{rt}$, see \cite{keyfitz1stedition} with $t=1$).
Similarly to the discrete formulation of rates of mortality improvement in \cite{Kannisto1994},
the minus sign in Equation \ref{estiromi} ensures that reductions in mortality result in positive values. \\

\begin{figure}[!ht]
\centering
\makebox{\includegraphics[height=0.7\textheight]{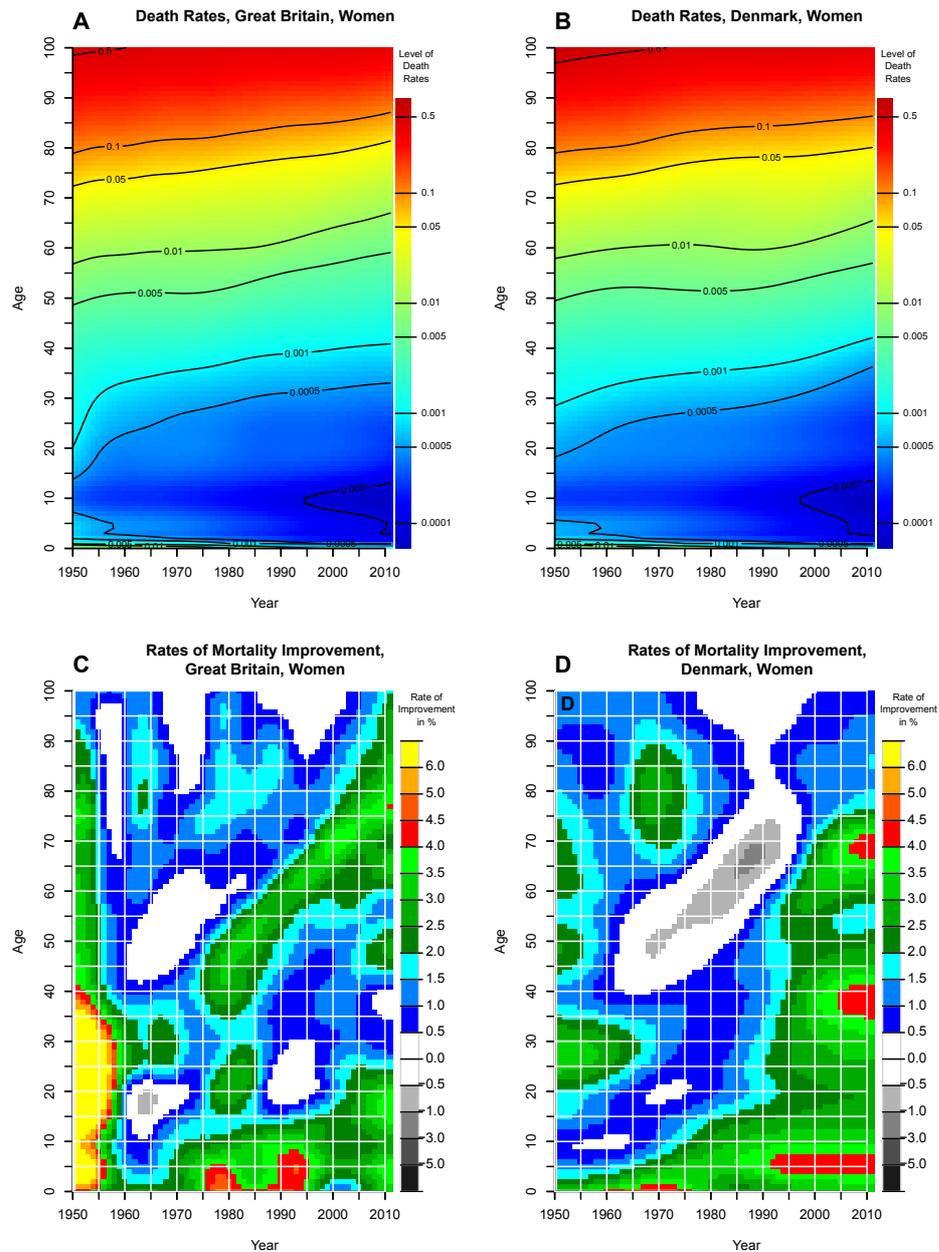}}
\caption{\label{mxandromi}Input of our model.
Upper Panel: (Smoothed) Death Rates for Women in Great Britain (A) and Denmark (B).
Lower Panel: Rates of Mortality Improvement (in \%) for Women in Great Britain (C) and Denmark (D).
Source: Authors' estimations based on data from the Human Mortality Database \citeyearpar{hmd2013}.}
\end{figure}

To illustrate why employing rates of mortality improvement
might be a better tool than age-specific death rates themselves,
Figure \ref{mxandromi} depicts in the upper two panels
for Great Britain (A) and Denmark (B) death rates on the so-called Lexis surface, i.\,e.~a plane by calendar time and age.
We smoothed the death rates with the $P$-spline smoothing methods of Currie et al. \citeyearpar{curriemortality},
based on \cite{Eilers1996} that are implemented in the R package \texttt{MortalitySmooth} of \cite{Camarda2012}.
Areas with the same color indicate the same level of mortality.
The aspect ratio of the two axes has been intentionally chosen that an increase of 10 years on the calendar time axis matches a 10 year increase on the age axis.
Thus, the trajectory of a birth cohort can be followed on a 45 degree line.
Contour lines were added to facilitate orientation on the surface.
It is evident in both pictures that a given color tends to reach higher ages over time, corresponding to a decrease in age-specific death rates for women in Great Britain and Denmark over time.
The antagonistic mortality dynamics in both countries, which we could expect from the diverging life expectancy trajectories, remain somewhat hidden in the two upper panels, though.
In our opinion, the display of rates of mortality improvement in the two lower panels provides better insights.
The various shades of grey indicate a negative improvement;
survival worsens in those areas.
Minor improvements are shown in blue, green colors were employed for moderate improvements.
Red, orange, and yellow were used for strong declines in mortality.
A 3.5\% annual decrease in mortality (a green level) translates, for example, to a cut in mortality by half in less than twenty years.
The divergent mortality dynamics in Great Britain and Denmark are clearly more visible in the surfaces of rates of mortality improvement (C and D):
While mortality decreased rather gradually in Great Britain in the last decades,
Danish women, born approximately between the two World Wars, experienced a strong cohort effect with stagnating or even deteriorating survival conditions during the 1980s and early 1990s,
which was the primary cause for the decelerated increase in Danish life expectancy in that period.

\subsection{Core Models}

We have two core models, a linear Bayesian hierarchical model and a Bayesian log-log model, that we can optionally use to forecast the development of longevity.
The advantage of our Bayesian core models is that they automatically model coherence of mortality change among adjacent ages to ensure,
for instance, a similar mortality decline for neighboring ages over time and a gradual increase in mortality in adult ages in each year. 
Another advantage is that they capture and quantify forecasting uncertainty,
i.\,e.~they give information about the spread and likelihood of our outcome. 

\subsubsection{Linear Bayesian model}
\label{sec:modelEquations}

One of our core models is a linear Bayesian hierarchical model.
A Bayesian hierarchical model can combine within-group and between-group information automatically \citep{Carlin2008, Gelman2006, Gelman2003, Jackman2009, Kruschke2011}.
In the context of mortality forecasting this means that such a model can capture the change of mortality within a certain age group as well as between adjacent age groups over time.
The level of heterogeneity in mortality across ages determines to what extent they will converge to an overall mean or a similar level.
The more mortality differs between adjacent ages over time, the less will the forecasted mortality levels converge to a similar level,
and the more will they follow their own trajectory.
In contrast, the less mortality differs between adjacent ages over time, the more will the forecasted mortality levels converge to a similar level,
and the less will they follow their own trajectory. \\   

In the linear Bayesian model, we can forecast the rates of mortality improvement $\rho$ by single age $x$ and year $t$ with a two-level normal model: \\

\begin{eqnarray}
\rho_{x,t} & \sim & N(\beta_{1,x} + \beta_{2,x}t, \sigma^2) \label{eq:lin1} \\
\beta_{j,x} & \sim & N(\mu_j,\tau) \label{eq:lin2} \\
\tau & \leftarrow & inverse(\Omega) \label{eq:lin3} \\
\Omega & \leftarrow & \bigl( \begin{smallmatrix}
                      \omega_1^2 & \rho\omega_1\omega_2\\
		      \rho\omega_1\omega_2 & \omega_2^2
                     \end{smallmatrix} \bigr ) \label{eq:lin4to7} \\
\sigma & \sim & U(0,1) \label{eq:lin8} \\
\mu_1 & \sim & U(-0.1,0.1) \label{eq:lin9} \\  
\mu_2 & \sim & U(-0.1,0.1) \label{eq:lin10} \\ 
\omega_1 & \sim & U(-0.1,1) \label{eq:lin11} \\ 
\omega_2 & \sim & U(-0.1,1) \label{eq:lin12} \\ 
\rho & \sim & U(-1,1) \label{eq:lin13} 
\end{eqnarray}

\begin{figure}[!ht]
\centering
\makebox{\includegraphics[width=0.7\textwidth]{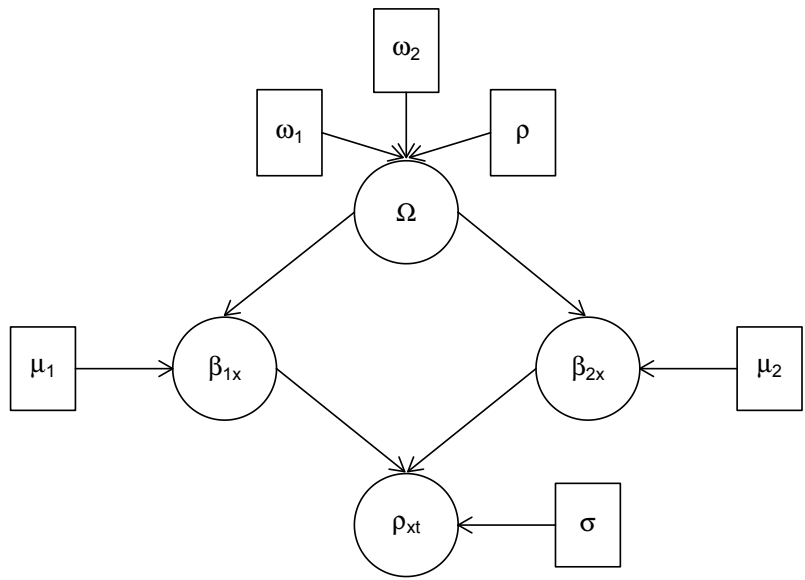}}
\caption{\label{fig:DAGLinearBHM} Directed acyclic graph for the linear Bayesian hierarchical regression model.
We apply this model to forecast rates of mortality improvement by single age and time.}
\end{figure}

In the first part, namely equation \ref{eq:lin1}, we model the trajectory for each age-specific rate of mortality improvement over time applying
a linear model with age-specific intercepts $\beta_{1,x}$ and age-specific slopes $\beta_{2,x}$. 
In the second part of our hierarchical model, namely equation \ref{eq:lin2}, we model the dependency of the rates of mortality improvement between adjacent ages $x$ by generating
overall means: $\mu_1$ for the $\beta_{1,x}$ and $\mu_2$ for the $\beta_{2,x}$.
Hence, combining information about mortality change within a certain age (part one) and between adjacent ages (part two) in one model
enables us to model a comprehensive picture of mortality improvement over age \emph{and} time.
To capture uncertainty,
we assume normally distributed realisations around mean rates of mortality improvement with a within age group variance $\sigma^2$ (equation \ref{eq:lin1})
and a between age group variance $\tau$ (equations \ref{eq:lin2} and \ref{eq:lin3}).
Moreover, we model a correlation structure between the intercepts and the slopes via the covariance matrix $\Omega$ (equation \ref{eq:lin4to7}).
$\omega_1$ is the standard deviation for the intercepts,
$\omega_2$ is the standard deviation for the slopes,
and $\rho$ is the correlation between the intercepts and the slopes across all ages $x$.
In equations \ref{eq:lin8} to \ref{eq:lin13}, we set vague priors for the hyperparameters $\sigma$, $\mu_1$, $\mu_2$, $\omega_1$, $\omega_2$, and $\rho$.
Figure \ref{fig:DAGLinearBHM} depicts the directed acyclic graph of our Bayesian hierarchical regression model. \\

\subsubsection{Bayesian log-log model}

In case we do not expect the rates of mortality improvement to follow a linear trend ad infinitum,
we can use the log-log model instead of the linear model.
The Bayesian log-log model uses logarithmic transformations to model non-linear survival improvements that can approach long-term levels (minimum, maximum) due to dampening reductions in mortality.
This also enables the Bayesian log-log model to capture the rising uncertainty of future mortality trends more realistically than the Bayesian linear model,
particularly in the long-run.\\

We re-estimate the trajectory of each age-specific rate of mortality improvement $\rho_x$ for each year $t$ in the base period on the logarithmic scale with a linear model in \emph{R} \citep{R2012}:

\begin{eqnarray}
\log(\rho_{x,t}) & = & \theta_{1,x} + \theta_{2,x} \cdot \log(t)
\end{eqnarray}

The linear model function uses the least square regression method to find a set of coefficients that resemble the observed data most so that there is a minimum of residuals between observed and predicted values.
We define the first year of the base period to be the origin of time,
assuring that our model obtains parameter estimates that are insensitive to the year of which mortality input data come from.
Without this definition, our model would generate different estimates for the same trajectory of $\rho_{x,t}$,
solely to the fact that they originate from different time periods since, e.\,g., $\log(t = 1750)$ does not equal $\log(t = 1950)$.
We can then take the estimated coefficients, i.\,e.~the intercepts $\theta_{1,x}$ and the slopes $\theta_{2,x}$, to compute the predicted rates of mortality improvement $\widehat{\rho}_{x,t}$:

\begin{eqnarray}
\widehat{\rho}_{x,t} & = & \exp(\theta_{1,x} + \theta_{2,x} \cdot \log(t))
\end{eqnarray}

The second coefficient $\theta_{2,x}$ is the constant elasticity of the rate of mortality improvement with respect to time,
i.\,e.~it gives information about the change of the rate of mortality improvement given a one percent change in time \citep{Hyndman2012}. \\ 

We use both coefficients as additional input parameters in our Bayesian log-log model:   

\begin{eqnarray}
\rho_{x,t} & \sim & N(\exp(\beta_{1,x} + \beta_{2,x} \cdot \log(t)), \sigma^2) \\
\sigma & \sim & U(0,1) \\
\beta_{1,x} & \sim & N(\theta_{1,x},\sigma_1^2) \\
\beta_{2,x} & \sim & N(\theta_{2,x},\sigma_2^2) \\
\sigma_1 & \sim & U(0,1) \\  
\sigma_2 & \sim & U(0,1) 
\end{eqnarray}

\begin{figure}[!ht]
\centering
\makebox{\includegraphics[width=0.7\textwidth]{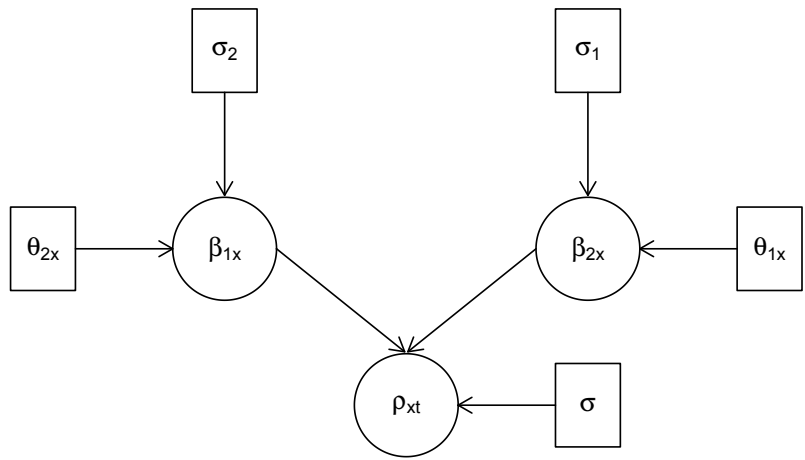}}
\caption{\label{fig:DAGLogBM}Directed acyclic graph for the Bayesian log-log model.
We apply this model to forecast rates of mortality improvement by single age and time.}
\end{figure}

In the Bayesian log-log model, we forecast the trajectory for each age-specific rate of mortality improvement $\rho_{x,t}$ over time
using an exponential function with log-transformed variables and with normally distributed deviances.
The exponential function uses age-specific intercepts $\beta_{1,x}$ and slopes $\beta_{2,x}$ as well as log-transformed time $log(t)$.
We use the point estimates for the coefficients $\theta_{1,x}$ and $\theta_{2,x}$ from the linear model function as marginal medians for the $\beta_{1,x}$ and $\beta_{2,x}$ and
we set vague priors for all standard deviations, i.\,e.~for $\sigma$, $\sigma_1$ and $\sigma_2$, that all represent uncertainty in the observed mortality data.
Figure \ref{fig:DAGLogBM} shows the directed acyclic graph for the Bayesian log-log model. 

\subsubsection{Implementation} 

Our goal is to find the unique and invariant forecasting (or posterior) distribution for the rates of mortality improvement with our core models using simulation-based Bayesian inference,
i.\,e.~we use the Gibbs sample algorithm \citep{Geman1984} to explore the forecasting distribution simulatively.\\

To compute and analyze the forecasting distribution, we use the statistical software \emph{R} \citep{R2012},
and \emph{Just Another Gibbs Sampler (JAGS)}, a freely available program that can be deployed for Bayesian analysis \citep{Plummer2011}.
We use the \emph{R} packages \emph{rjags} \citep{Plummer2013} and \emph{R2jags} \citep{Su2013} to interface between \emph{R} and \emph{JAGS}. \\

\emph{JAGS} requires certain input data to compute the forecasting distribution for the rates of mortality improvement with our core models.
Hence, we have to define the equations of our core models in a specific programming language for \emph{JAGS}.
In addition, we have to provide basic data in \emph{R} that will be used from \emph{JAGS} to execute the models.  
To this basic data belong, for instance, the observed rates of mortality improvement, the number of single ages, the number of forecast years as well as the age-specific coefficients $\theta_{1,x}$ and $\theta_{2,x}$ for the log-log model.
Next to this basic data, we have to determine several execution properties like the random number generator, the number of parallel chains,
the number of iterations, the number of thinning as well as the length of the burn-in period. 
In general, it can be helpful to fit the forecasting models with a large number of chains and iterations to explore the parameter space exhaustively.
We use the R-package \emph{R2jags} to execute parallel chains with \emph{JAGS}.
Running parallel chains can substantially speed up the execution of many iterations.
Moreover, thinning, by means of saving only each \emph{j}-th iteration, can be used to avoid dependencies between adjacent iterations (autocorrelation).
For outcome analysis we only use iterations from the chain that has converged to the forecasting distribution.
Hence, we can use the burn-in period to identify and exclude initial parameter values that still converge to a constant mean from outcome analysis \citep{King2012}. \\

Running the Gibbs sampling algorithm, we obtain all model parameter estimates simultaneously and we forecast the mortality trends and their uncertainty coherently.
Hence, both of our core models, the linear one and the log-log one, provide probabilistic mortality forecasts. 

\subsection{Model Output}

The output of our mortality forecasting model are simulated posterior distributions for the future rates of mortality improvement.
These posterior distributions comprise all executed iterations;
for instance, when we run the Gibbs sampling algorithm with a total of $5,000$ iterations,
we also obtain a total of $5,000$ potential future values for each single rate of mortality improvement.
We then quantify the uncertainty, which is related with forecasting survival improvements,
by using the quantiles $i$ of the forecasted rates of mortality improvement to forecast the respective quantiles $i$ of the death rates $m$ by age $x$ and time $y$:

\begin{eqnarray}
m^{i}_{x,y} & = & m^{i}_{x,y-1}(1 - \rho^{i}_{x,y})
\label{eq:forecast_log_romi}
\end{eqnarray}

These forecasted death rates can then be used for further mortality analysis, such as life table calculations.
To summarize mortality in a population, we use, for instance, the forecasted death rates to compute life expectancy at birth.

\subsection{Additional notes}
\label{sec:advantageCommonAppr}

\subsubsection{Rates of mortality improvement}

Common mortality forecasting models often lack the possibility to model a dynamic age pattern of mortality change such as older ages increasingly experiencing stronger survival improvements.
For instance, one of the most commonly used mortality forecasting approaches, the original Lee-Carter model,
uses only one schedule that determines the relative mortality progress between the ages for all forecast years,
i.\,e.~it determines how strong mortality will improve for each age in relation to all other ages.
This may cause substantial errors,
even when forecasting mortality only in the short and medium run.
Such shortcomings also apply, albeit to a lesser extent, to many extensions of the original Lee-Carter model,
despite additional components in the predictor structure to optionally account for cohort and extra period effects \citep{Mitchell2013}.
To address this issue, very recently developed approaches \citep{Haberman2012,Mitchell2013} use rates of improvement rather than death rates to forecast mortality.
We also use the rates of mortality improvement to capture dynamic shifts in mortality reduction from younger to higher ages.
Hence, instead of using only one fixed age schedule of mortality improvement for all forecast years,
our model allows mortality to change at each age \emph{and} in each year to forecast.

\subsubsection{Optionally complement mortality trends}

Generating accurate mortality forecasts is especially challenging for populations,
which undergo an irregular mortality development leading to a rather unstable increase in life expectancy.
Danish women experienced such an unsteady mortality development:
The increase in life expectancy at birth stalled during the 1980s and early 1990s
and is now catching up to international trends at an accelerated pace.
Most forecasting methods would underestimate this progress in Danish life expectancy in the 1990s by extrapolating the mortality improvement of the 1980s and early 1990s.
We can avoid this error in our proposed model due to the optional combination of objective and subjective information:
If we expect such an objective and purely data-driven extrapolation of the past mortality trend as insufficient or implausible,
we can complement it with the mortality trend of at least one reference country.
In the case of Denmark we chose to complement the minor extrapolated Danish mortality reduction with the faster extrapolated trend of neighboring Sweden.
Comparable to the approach by \cite{Li2005},
we expect that this combination results in a more plausible mortality forecast for Danish women,
basically assuming that Danish women successively approach the faster mortality decline of Swedish women.
How many and which reference countries are selected is due to expert judgment.
We recommend to take reference countries whose mortality conditions are expected to be similar to those in the country of interest in coming years.
In case we use information from reference countries,
we model a successive transition of mortality improvement trends between the country of interest and the reference countries:
In the short term, we put more weight on the trend of the country of interest, while, in the long term, we shift the weight towards the trends of the reference countries. 
This procedure is similar to methods proposed by \cite{Li2005} or \cite{Cairns2011},
which forecast mortality of multiple countries jointly.
But instead of setting the main focus on generating consistent mortality forecasts for \emph{n} populations,
we use mortality trends of reference countries to adjust a purely extrapolated mortality trend in a country of interest with additional sources of information like expert judgment.

\subsubsection{Capture forecast uncertainty}

Since we do not know with certainty how mortality will develop in the future,
we propose a probabilistic model,
which forecasts multiple potential scenarios together with their occurrence probability.
To capture and quantify the inherent uncertainty of mortality forecasts,
we use a Bayesian model that provides prediction intervals for each age-specific death rate in a given year.
In the standard version of our model,
we use unsmoothed death rates as input data to capture the natural variablity in mortality;
the width of the resulting prediction intervals is comparable to those of other approaches like the Lee-Carter model or the Bayesian model from the UN.
However, our model also allows to work with smoothed death rates.
To smooth the death rates, we use the R package \texttt{MortalitySmooth} by \cite{Camarda2012};
it implements the $P$-spline smoothing methods of \cite{curriemortality}, which use methods of \cite{Eilers1996} with a particular focus on mortality data \citep{Camarda2008}.
The consequence of smoothing, however, is,
while the median is almost indistinguishable from our standard model,
the prediction intervals are much smaller since we removed virtually all natural variability of the observed death rates.
Nonetheless, both approaches can be conducted---the one with unsmoothed death rates to capture the natural variability in the prediction intervals
and the other one with smoothed death rates if only point estimates are of interest.

\section{Application}
\label{ch:application}

In this section, we compare the performance of our proposed model with
several well-established models to project mortality.

An obvious choice is the original model by \cite{leecarter} due to its worldwide acceptance and application in the field of mortality forecasting \citep{Booth2006b, Shang2011}.
Numerous enhancements of the Lee-Carter model have been proposed since its introduction about twenty years ago.
For our comparison, we select three generalizations suggested by Renshaw and Haberman \citeyearpar{Renshaw2003, Renshaw2006} who add terms regarding cohort and extra period effects to the original predictor structure.

\cite{Li2005} argue that two or more countries with similar characteristics are very unlikely to follow divergent mortality trends in the long run.
Thus, in 2005, they introduced the concept of ``coherent mortality forecasting''.
This approach, which we also employ, assumes that ``populations of the world are becoming more closely linked by communication,
transportation, trade, technology, and disease'' (p.~575).
It would make sense, therefore, to model death rates changing at the same rate in comparable populations instead of conducting forecasts for individual countries.

Apart from the class of those Lee-Carter models, we also compare our model with the Bayesian approach of \cite{Raftery2013} that is used by the UN Population Division for the \emph{World Population Prospects 2012} \citep{UN2013}.
This approach forecasts life expectancy probabilistically with a Bayesian Hierarchical Model using a time-series approach with country-specific and overall country information (in a double-logistic function).

In addition, we also compare our approach to the one proposed by \cite{curriemortality}.
It is based on a two-dimensional, non-parametric smoothing approach using $P$-splines \citep{Eilers1996}.\\

All models are tested using data for Great Britain and Denmark.
We select these two countries because women in Great Britain feature a rather regular mortality development with a stable increase in life expectancy at birth.
Danish women, in contrast, experienced a period of virtual stagnation during the 1980s until the middle of the 1990s,
caused by the widespread prevalence of smoking of women born between the two World Wars \citep{jacobsen, panel2010}.
The detection of such structural changes in mortality trends has been analyzed recently \citep{Vallin2009, vanberkum2013} and poses additional obstacles to forecast mortality accurately.\\

We forecast mortality for Great Britain and Denmark with all these models in two settings:
a) a retrospective (or in-sample) forecast and b) a prospective (or out-of-sample) forecast.
The retrospective forecast projects mortality from 1991 to 2011, based on data from 1965 through 1990.
This allows us to compare the estimates of the models with the observed development.
Due to the regular trajectory of mortality development, we expect that {\em all} models perform reasonably well for Great Britain.
In the case of Denmark, we hypothesize that our model will outperform others due to (1) the flexible age pattern of mortality improvement over age and time
in combination with (2) the optional combination of mortality trends from the country of interest as well as from other reference countries.

In the second forecast setting,
we conduct a prospective forecast for both countries,
i.\,e.~we take death rates between 1965 and 2011 as base period data in order to forecast them from 2012 to 2050.
As we do not know how mortality will actually develop in the future, we cannot compute forecast errors by comparing forecasted and observed mortality data.
Alternatively, we compare our forecasts of life expectancies at birth with those of other methods as well as with published forecasts by agencies like Eurostat. 

\subsection{Data and parameter settings}

Death rates, death counts and/or population exposures,
which are respectively required by the models as input data,
were all obtained from the Human Mortality Database \citeyearpar{hmd2013}.

\subsubsection{Our model}

To forecast mortality with our model for women in Great Britain and Denmark,
we take death counts and exposure times for single ages and years from the Human Mortality Database \citeyearpar{hmd2013}.
For Great Britain we only consider mortality data from the country of interest,
whereas for Denmark we also consider mortality data from Sweden in the retrospective forecast
and, additionally, from France, Italy and Japan in the prospective forecast,
assuming that Denmark is likely to catch up with international trends again.
To capture and quantify forecast uncertainty,
we use unsmoothed mortality data,
i.\,e. we compute the rates of (mortality) improvement for each age-specific death rate over time.
We then forecast these rates of mortality improvement with our Bayesian log-log core model via \emph{JAGS} by letting the Gibbs sample algorithm run with five parallel chains for a total of $5,200$ iterations after a burn-in period of $200$ iterations.
To avoid dependence between adjacent trials, we only save each fifth iterate for inferences.
In addition, we exclude implausibly low and high forecasts for the rates of mortality improvement by setting thresholds,
i.\,e.~they can neither fall below $0.005$ nor can they exceed $0.035$.
In case the model approaches minimum or maximum rates of mortality improvement,
the forecasted death rates still decline with time.
Table \ref{tab01} summarizes all parameter settings for our mortality forecasts.

\begin{table}
\caption{\label{tab01}Model parameters and their values in the mortality forecasts for British and Danish women.}
\centering
\fbox{
\begin{tabular}{*{2}{c}}
Model parameter & Parameter values \\
\hline
Country of interest & Great Britain/Denmark \\
Reference country (retrospective) & Great Britain/Sweden \\
Reference country (prospective) & Great Britain/Sweden, France, Italy, Japan \\
Sex & Females \\
Minimum age & $0$ \\ 
Maximum age & $110+$ \\
Base period (retrospective) & $1965-1990$ \\
Base period (prospective) & $1965-2011$ \\
Forecast horizon (retrospective) & $21$ years ($1991-2011$) \\
Forecast horizon (prospective) & $39$ years ($2012-2050$) \\
Bayesian core model & log-log \\ 
Number of iterations & $5,200$ \\
Number of adaptions & $200$ \\
Number of parallel chains & $5$ \\
Number of thinning & $5$ \\
Adjust forecasted $\rho_{x,t}$ & TRUE (min: $0.005$, max: $0.035$) \\
\end{tabular}}
\end{table}

\subsubsection{Lee-Carter Model}

The mortality forecasts with the Lee-Carter model are estimated with the freely available implementation in \emph{R} by Timothy \citeauthor{Miller},
which requires death rates by age and time.

\subsubsection{Lee-Carter Extensions by Renshaw \& Haberman}

We generate the mortality forecasts with three Lee-Carter extensions proposed by Renshaw and Haberman \citeyearpar{Renshaw2003,Renshaw2006},
applying the \emph{R}-packages \emph{ilc} \citep{Butt2010b} and \emph{demography} \citep{demography}.
We fit the three models, labeled as \emph{h0, h1,} and \emph{h2} by the original authors \citep{Renshaw2003,Renshaw2006},
with the \emph{lca.rh} routine that needs death rates and exposures, arranged in a \emph{demogdata} object, as input.
For most parameters, we use the default settings.
Exceptions are parameters determining the type of the model and of the error structure,
which we set to the respective model (h0, h1 or h2) and to \emph{gaussian} or \emph{poisson};
in addition, we set the \emph{spars-}~parameter to the recommended value of 0.6,
which is intended to smooth the input data.
To forecast the fitted models,
we use the \emph{forecast-}~function,
setting the \emph{jump.choice-}~parameter to \emph{actual},
which produces more plausible mortality forecasts than setting it to \emph{fit}.
The output contains forecasted death rates and life expectancies at birth for the respective forecast years.
In our comparative analysis, we only take the parameter settings for each of these three models, which achieve the best forecasting results. 

\subsubsection{\emph{P}-Spline Approach}

The $P$-spline approach to smooth and forecast mortality over age and time, introduced by \cite{curriemortality},
models death counts with the natural logarithm of the exposed population as an offset in a Poisson setting.
We employ the software implementation by \cite{Camarda2012}.
Two parameters are of crucial importance: the smoothing parameter(s) $\lambda$ and the order of the penalty.
The $\lambda$s are found by optimizing the BIC on a large grid of potential $\lambda$ values.
Specifying the correct order of the penalty is less straightforward, as pointed out in the original paper \citep[p.~289, 297]{curriemortality}.
An order of the penalty of 1 refers to future mortality at a constant level, 2 to improvements at a constant rate and 3 to an accelerating rate.
After experimentation, we choose the default order of the penalty of 2, which gives the best results in our retrospective framework.

\subsubsection{Coherent Lee-Carter}

Following the suggestion by Ronald D. Lee (personal communication),
we estimated the ``Coherent Lee Carter'' model by \cite{Li2005} with the web-based platform LCFIT,
hosted on the website of Berkeley's demography department.
The interface requires mortality rates for at least two countries and, optionally, population counts.
We used the default settings of the optional parameters after experimentation has shown that they produce the best results.
To generate plausible mortality forecasts with the coherent Lee-Carter model,
we supplemented the Danish data with death rates and exposures from Sweden in the retrospective forecast
and, additionally, with mortality data from France, Italy and Japan in the prospective forecast.

\subsubsection{Bayesian approach by Raftery et al.}

We use the freely available R-package \emph{bayesLife} \citep{Sevcikova2013} to forecast mortality with the Bayesian approach by \cite{Raftery2013}.
Applying the \emph{run.e0.mcmc} and the \emph{e0.predict} routine with default parameter values,
we simulate 160,000 iterations with 3 chains.
The exploration of the trajectories for future life expectancy yields quantiles only for quinquennial data,
i.\,e.~in the retrospective forecasts this approach provides life expectancy for the five year periods 1990-1995, 1995-2000, 2000-2005 and 2005-2010
of which we take the mid points 1993, 1998, 2003 and 2008. 

\subsection{Retrospective forecast}

\begin{figure}[!ht]
\centering
\makebox{\includegraphics[width=1\textwidth]{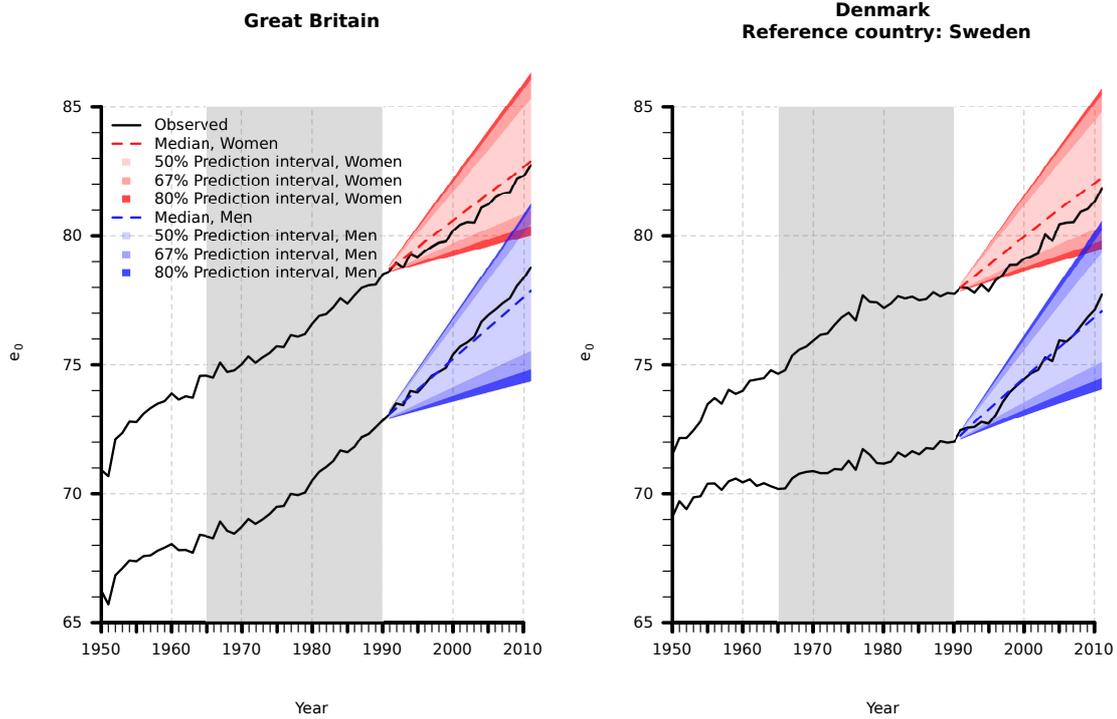}}
\caption{\label{fig:Figure_5a_BohkRau}Observed (black) and forecasted life expectancy ($e_0$) of our model for British (left) and Danish (right) women (red) and men (blue);
increasing forecast uncertainty is represented by the widening 80\%, 67\%, and 50\% prediction intervals,
whose colors become brighter from the outside to the median (dashed line).
In this in-sample forecast, we take the data from 1965 to 1990 (gray colored box) as basis to forecast mortality from 1991 to 2011.
Moreover, we complement the mortality trend for Danish women with that of Swedish women.}
\end{figure}

\paragraph{Prediction intervals}In our probabilistic mortality forecasts from 1991 to 2011,
we capture and quantify the natural variability (or the inherent uncertainty) using raw mortality data as input.
Figure \ref{fig:Figure_5a_BohkRau} depicts the median and the 50\%, 67\% and 80\% prediction intervals for life expectancy at birth for both sexes in Great Britain and Denmark:
Our model forecasts increasing life expectancy with the observed values fluctuating (narrowly) around the median.
Moreover, the forecast uncertainty increases with time---an effect that is represented by gradually widening prediction intervals.
For instance, with a probability of 80\%, life expectancy of British women will range between 79 years and 82 years in 2000,
whereas it will range between 80 years and 86 years in 2011.
Hence, the (absolute) width of the 80\% prediction interval becomes twice as large between the tenth and the twenty-first forecast year,
i.\,e. it increases from three to six years.
That the width of the prediction intervals of our model is comparable to those of the original Lee-Carter model,
of the $P$-Spline approach, of the LC-coherent model and of the UN Bayesian approach is illustrated for British and Danish women in Figure \ref{fig:PredictionIntervals_DiffApproaches_DNKGBR_1965_1990_2011} in the appendix.
In 2011, the width of the 95\% prediction intervals ranges between 4.6 years and 6.5 years among these models.

\begin{figure}[!ht]
\centering
\makebox{\includegraphics[width=0.75\textwidth]{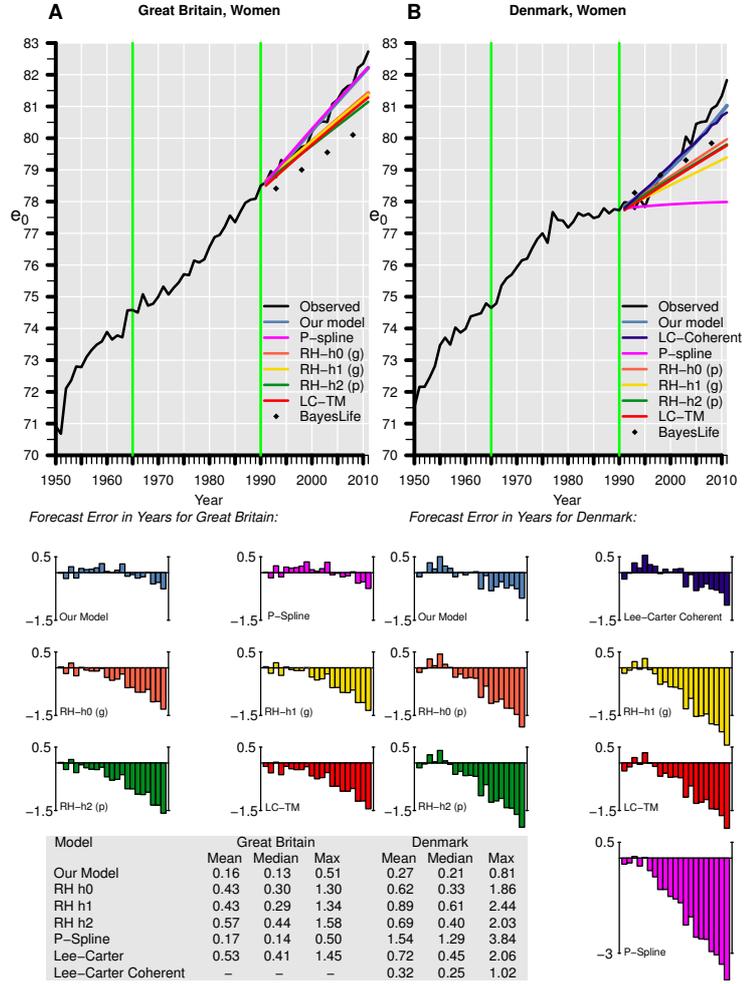}}
\caption{\label{fig:BohkRau_GBR_loglog_LC_inSample_e0.pdf}Observed (black) and forecasted life expectancy ($e_0$) of our model (blue),
of the $P$-spline approach (magenta), of the original Lee-Carter model (red) and of three of its refinements ($h0$ (light red), $h1$ (yellow),
and $h2$ (green)), of the coherent Lee-Carter model (navy blue) and of the UN Bayesian approach (black squares) for women in Great Britain (A) and Denmark (B). 
In this retrospective forecast from 1991 to 2011, we take the data from 1965 to 1990 (green vertical reference lines) as basis.
Moreover, we complement the mortality trend for Danish women with that of Swedish women in our model as well as in the coherent Lee-Carter forecast.
In the Renshaw-Haberman models, (p) and (g) denote poisson and gaussian errors, respectively.
Our model (blue) has substantially smaller forecast errors for both populations;
it deviates less from observed values than the other models.}
\end{figure}

\paragraph{Median forecasts}
We use forecasts of the median for the years 1991 to 2011 to compare the forecasting performance of all selected approaches.
Figure \ref{fig:BohkRau_GBR_loglog_LC_inSample_e0.pdf} depicts the observed and forecasted life expectancy at birth for all models.
The green vertical reference lines indicate the beginning (1965) and the end (1990) of the period used as the base for the retrospective forecasts from 1991 to 2011.\\

The forecast errors, defined as the difference between forecasted and actually observed life expectancies at birth $e_0$ from 1991 to 2011:

\begin{eqnarray}
E_t & = & e_{0,t}^{forecast} - e_{0,t}^{observed}
\label{eq:forecast_error}
\end{eqnarray}

are displayed for each model below the two panels.
In the case of Great Britain, the $P$-spline model as well as our model provide a satisfactory fit.
The mean absolute errors of the models are 0.17 years and 0.16 years, respectively.
Considering that (record) life expectancy increases annually at a pace of about 0.25 years, the deviation is negligibly small.
Whereas the errors in these two models are not systematically biased,
the errors in the original Lee-Carter model as well as the errors in its extensions by Renshaw and Haberman accumulate.
The mean absolute error in those four models ranges between 0.43 and 0.57 years;
life expectancy in 2011, the end of our retrospective forecasting period, is underestimated by 1.30 years to 1.58 years.
Using the web-based interface LCFIT instead of Timothy Miller's code for the original Lee-Carter model does not yield a better estimate.

As pointed out previously: Denmark's life expectancy (right panel in Figure \ref{fig:BohkRau_GBR_loglog_LC_inSample_e0.pdf}) is rather unusual
with a period of virtual stagnation during the 1980s and early 1990s,
followed by a period of catching up to international trends.
Surprisingly, the nonparametric $P$-spline approach produced the largest forecast errors.
Neither modifying the order of the penalty, corresponding to different assumptions of the trajectory of future mortality \citep{curriemortality},
nor adding weights, which change over time, as provided as an option in the package \texttt{MortalitySmooth} \citep{Camarda2012}, improves the $P$-spline forecast.
The original Lee-Carter model and its modifications by Renshaw and Haberman perform better.
Figure~\ref{fig:BohkRau_GBR_loglog_LC_inSample_e0.pdf} illustrates that taking a reference population (i.e.~Sweden) into account is the key
to estimate mortality and life expectancy if it is expected that an extrapolation of past trends in a single country of interest appears improbable.
The {\em maximum} absolute error of the ``coherent forecast'' by \cite{Li2005} is 1.02 years.
Our model performs even slightly better with a maximum absolute error of 0.81 years.
Thus, the {\em maximum} absolute error in our model is smaller than the {\em mean} absolute error in the $h1$ model (0.89 years)
and only marginally worse than the mean absolute error of the original Lee-Carter model and the other two extensions by Renshaw and Haberman (ranging between 0.62 years and 0.72 years).

We assume that the way {\em how} the trend of the reference country is incorporated into the model explains the smaller errors of our model:
While the coherent Lee-Carter framework assumes a joint trend of the countries,
our approach models a declining path dependence of the extrapolated trend for the country of interest (Denmark) jointly with an increasing importance of the trend of the reference country (Sweden).
Specifically in our application:
In 1991, the first forecast year, we use 100\% of the extrapolated trend of Denmark and 0\% of the extrapolated trend of Sweden.
Denmark's weight decreases linearly over time until it is non-existent in 2011,
whereas Sweden gains importance during the forecast period, dominating completely the forecast in 2011. 

\subsection{Prospective forecast}

\begin{figure}[!ht]
\centering
\makebox{\includegraphics[width=1\textwidth]{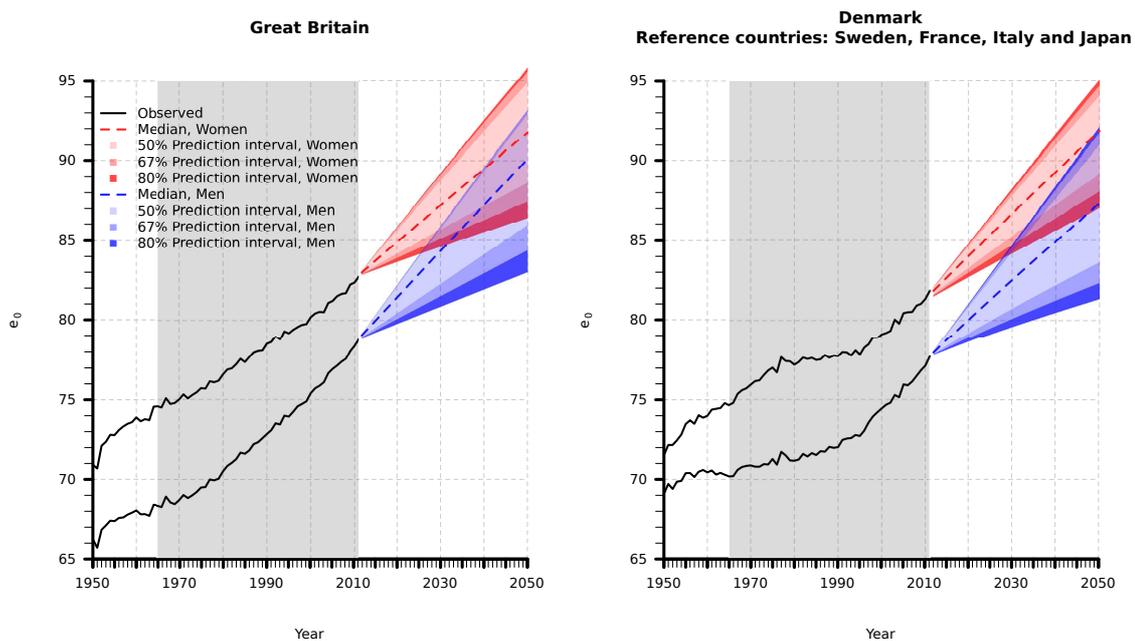}}
\caption{\label{fig:Figure_6a_BohkRau}Observed (black) and forecasted life expectancy ($e_0$) of our model for British (left) and Danish (right) women (red) and men (blue);
increasing forecast uncertainty is represented by the widening 80\%, 67\%, and 50\% prediction intervals,
whose colors become brighter from the outside to the median (dashed line). 
In this prospective forecast, we take the data from 1965 to 2011 (gray colored box) as basis to forecast mortality from 2012 to 2050.
Moreover, we complement the mortality trend for Danish women with that of Swedish, French, Italian and Japanese women.}
\end{figure}

\paragraph{Prediction intervals}In our probabilistic mortality forecasts from 2012 to 2050,
we capture and quantify increasing life expectancy and its uncertainty with the median and gradually widening 80\%, 67\%, and 50\% prediction intervals
that are illustrated in Figure \ref{fig:Figure_6a_BohkRau} for women and men in Great Britain and Denmark.
In 2050, female life expectancy ranges between 86.39 years and 95.81 years in Great Britain,
and between 87.02 years and 95.06 years in Denmark with a probability of 80\%.
It is remarkable that the difference between the lower quantiles is larger than the difference between the upper quantiles,
indicating that the uncertainty (or the variability) is greater for slower increases than for stronger increases in life expectancy in these two countries.

\begin{figure}[!ht]
\centering
\makebox{\includegraphics[width=1\textwidth]{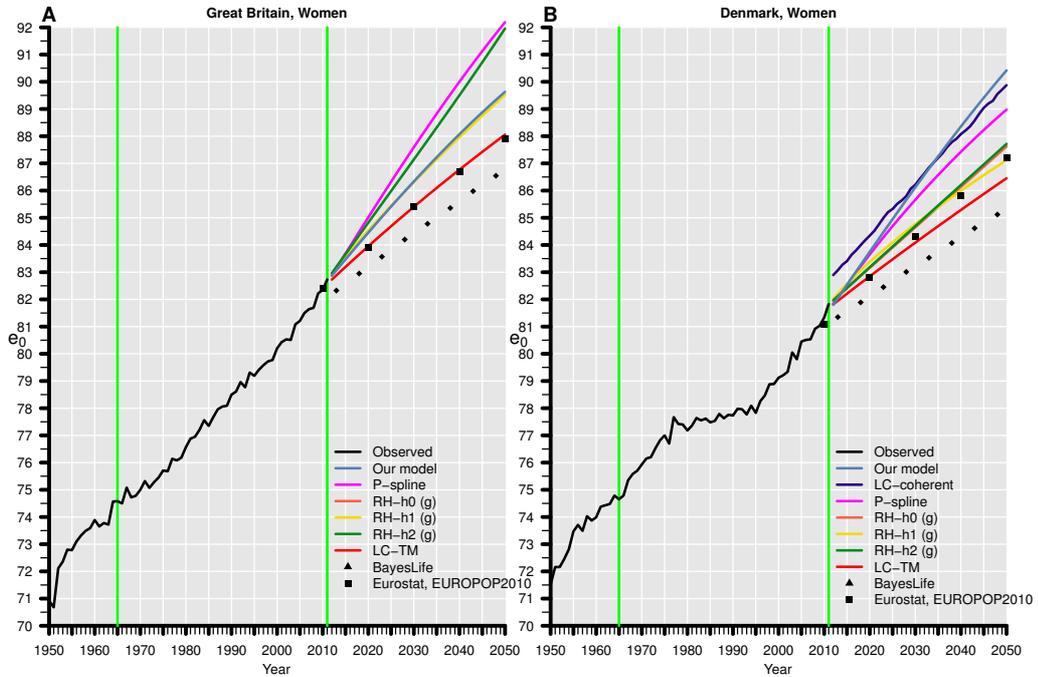}}
\caption{\label{fig:BohkRau_DNK_GBR_loglog_LC_EUR_UN_outOfSample_e0}Observed (black line) and forecasted life expectancy ($e_0$) of our model (blue),
of the $P$-spline approach (magenta), of the original Lee-Carter model (red) and of its three refinements $h0$ (light red), $h1$ (yellow),
and $h2$ (green), of the coherent Lee-Carter model (navy blue), of the UN Bayesian approach (black triangles) and of Eurostat (black squares)
for women in Great Britain (A) and Denmark (B). 
In this prospective forecast from 2012 to 2050, we take the data from 1965 to 2011 (green vertical reference lines) as basis.
Moreover, we complement the mortality trend for Danish women with that of Swedish, French, Italian and Japanese women in the coherent Lee-Carter forecast as well as in our model.
In the Renshaw-Haberman models, (p) and (g) denote poisson and gaussian errors, respectively.} 
\end{figure}

\paragraph{Median forecasts}In Figure \ref{fig:BohkRau_DNK_GBR_loglog_LC_EUR_UN_outOfSample_e0},
we compare the median prospective forecasts of life expectancy from 2012 to 2050 of our model,
of the original Lee-Carter model and its extensions $h0$, $h1$, and $h2$, of the coherent Lee-Carter model, of the $P$-spline approach,
of the UN Bayesian approach and of the Eurostat \emph{EUROPOP2010} forecast \citep{EC2011} for British and Danish women.
For each model fit, we take the death rates from 1965 to 2011 in the British and Danish mortality forecast. \\

Comparing these prospective mortality forecasts for British and Danish women reveals several aspects.
First, all models predict for both countries the almost linear increase in female life expectancy of the recent past to continue,
albeit the models applying a Lee-Carter predictor structure forecast a larger progress for British than for Danish women.
For instance, they predict an average increase in life expectancy of approximately 7 years for British women and of 5 years for Danish women between 2012 and 2050.
This smaller progress in Danish life expectancy is probably due to the interrupted linear increase during the 1980s and early 1990s.
Together with the continous linear increase in British life expectancy, this clearly demonstrates that selecting the base period with a certain mortality trend has a decisive influence on forecast outcome \citep{vanberkum2013},
in particular for purely data-driven methods that extrapolate past trends.
Since we expect Danish females to catch up to international trends again,
we supplement the Danish mortality trend with that of Swedish, French, Italian and Japanese women in our model as well as in the coherent Lee-Carter model.
As a consequence, both models forecast (in comparison to the other applied models) the strongest increases in Danish female life expectancy,
i.\,e. our model forecasts an increase of 8.6 years between 2012 and 2050
and the coherent Lee-Carter model forecasts an increase of 7 years in the same period.
Although both approaches jointly forecast multiple mortality trends,
the trajectory of our model continues the long-time trend more plausibly than the coherent Lee-Carter model,
which lacks a smooth transition between the observed and forecasted life expectancy---at least we were unable to remove this discrepancy using the web-based platform LCFIT.
Second, given the large variation in the forecasts of British life expectancy among all approaches,
the results of our model are rather in the center than near the lower or upper bound of all forecasts.
A different picture emerges for Denmark:
Despite the somewhat smaller variation in the mortality forecasts among all approaches,
our model forecasts relatively large gains in life expectancy compared to the UN Bayesian approach or to the original Lee-Carter model,
which provide rather conservative forecasts.
Third, despite all predicting almost linear trends,
forecasts for life expectancy vary considerably among the models.
For instance, the $P$-spline approach forecasts mortality with one of the highest life expectancy values in 2050:
For British women, it is 92.18 years and for Danish women, it is 88.97 years.
In contrast, the original Lee-Carter model forecasts mortality for both populations with relatively low life expectancy in 2050:
For British women, it is 88.05 years and for Danish women, it is 86.45 years.
Only the UN Bayesian approach provides forecasts with even smaller life expectancy than the original Lee-Carter model:
For British women, it is 86.54 years and for Danish women, it is 85.12 years (in 2048).
Fourth, each of these forecasting models provides prediction intervals to capture forecast uncertainty.
For instance, our model as well as the $P$-spline approach generate relatively wide prediction intervals:
In 2050, the life expectancy of British women is estimated to range between 85 years and 96 years (according to our model) or to range between 87 years and 96 years (according to the $P$-spline approach) with a probability of 95 percent.
In contrast, the other approaches provide smaller prediction intervals;
for instance, the original Lee-Carter model predicts British female life expectancy to range between 85 years and 90 years with a probability of 95 percent. \\

\begin{figure}[!ht]
\centering
\makebox{\includegraphics[height=0.7\textheight]{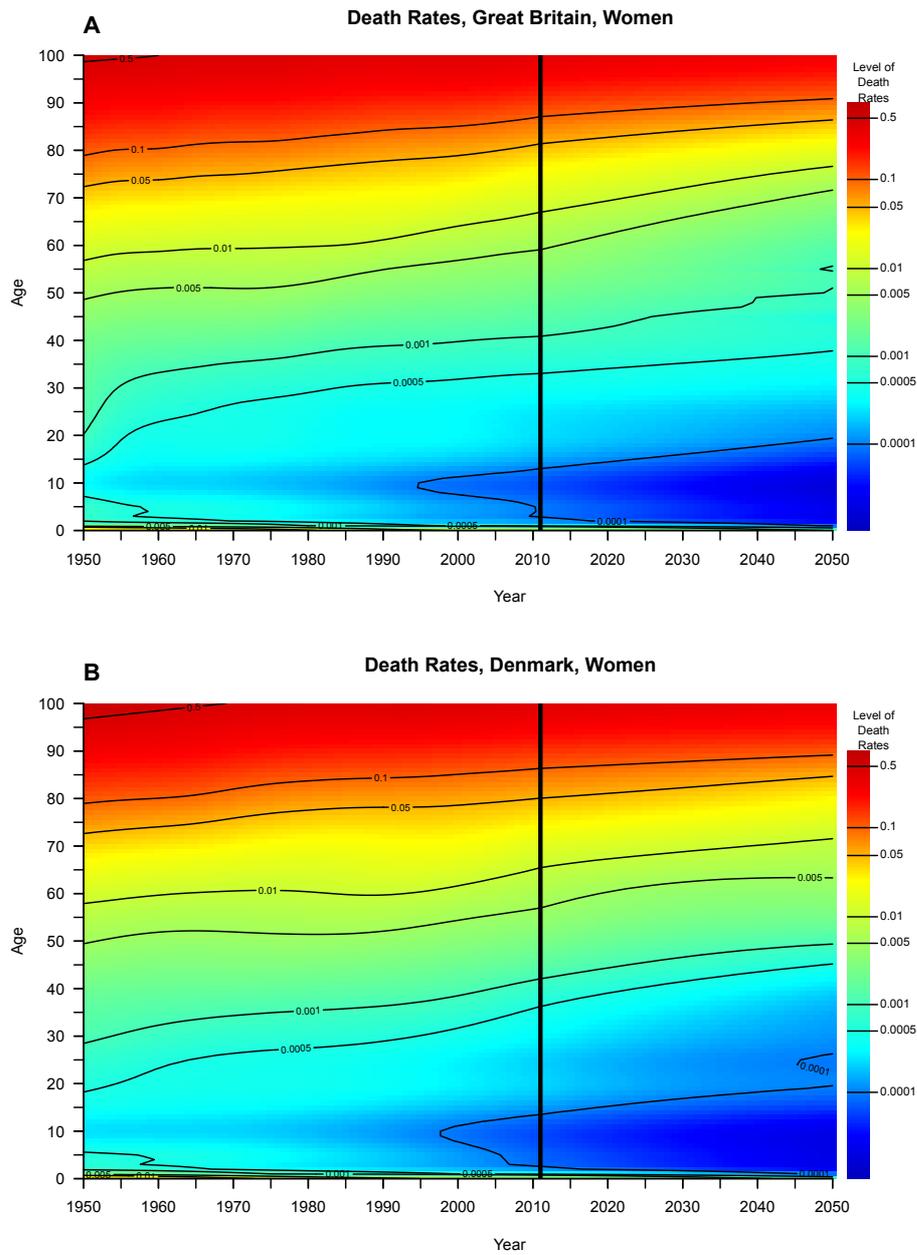}}
\caption{\label{fig:SurfacesDeathRates1950.2050GBRandDNKwomen}Death rates from 1950 to 2050 for British and Danish women.
Death rates by single age observed from 1950 to 2011 and forecasted with our proposed model from 2012 to 2050 for women in Great Britain (A) and Denmark (B). 
The transition from observed to forecasted death rates is illustrated by the vertical black reference line in both displays.}    
\end{figure}

The basis for those forecasted life expectancies at birth are the forecasted death rates
that are shown for the base and forecast years of our prospective forecasts in Figure \ref{fig:SurfacesDeathRates1950.2050GBRandDNKwomen}. 
A closer look at these two panels, based on smoothed input data, reveals at least two things.
First, the transition between the observed and forecasted death rates is smooth:
There are visually no distortions between the base and the forecast period,
which are separated by a vertical reference line in the year 2011.
Second, the death rates decline gradually for each age over time
as reflected by the fine color gradient for each age as well as by the slightly increasing contour lines.
Hence, we avoid erratic and spurious death rates in our forecasts as illustrated by a continous rather than a noncontinous color flow for each age with time. 
Additionally, the slightly increasing contour lines indicate that certain mortality levels reach successively higher ages,
i.\,e.~mortality reduces gradually over time without erratic fluctuations.
Furthermore, the plots illustrate that our model does not generate implausible estimates for the forecasted age pattern of mortality;
for instance, mortality at age 81 never drops below the level at age 80 in a given year. 

Although, we only show the outcome for the mortality forecasts of our log-log model her,
the outcome of our linear model is similar.
We illustrate the close resemblance of both approaches in Figure~\ref{fig:SurfacesLinearDeathRates1950.2050GBRandDNKwomen} in the Appendix,
which plots observed and estimated death rates from the linear model.
The results for the estimated death rates are similar to those from the log-log model, displayed in Figure~\ref{fig:SurfacesDeathRates1950.2050GBRandDNKwomen}. 

\subsection{Diagnostics}

Next to the projection outcome, we also have to check the model performance of our model,
i.\,e.~we assess its convergence via trace plots and autocorrelation functions for the model parameters,
and we control its run-length via the Raftery-Lewis diagnostic \citep{Raftery1992}. \\ 

\begin{figure}[!ht]
\centering
\makebox{\includegraphics[width=1\textwidth]{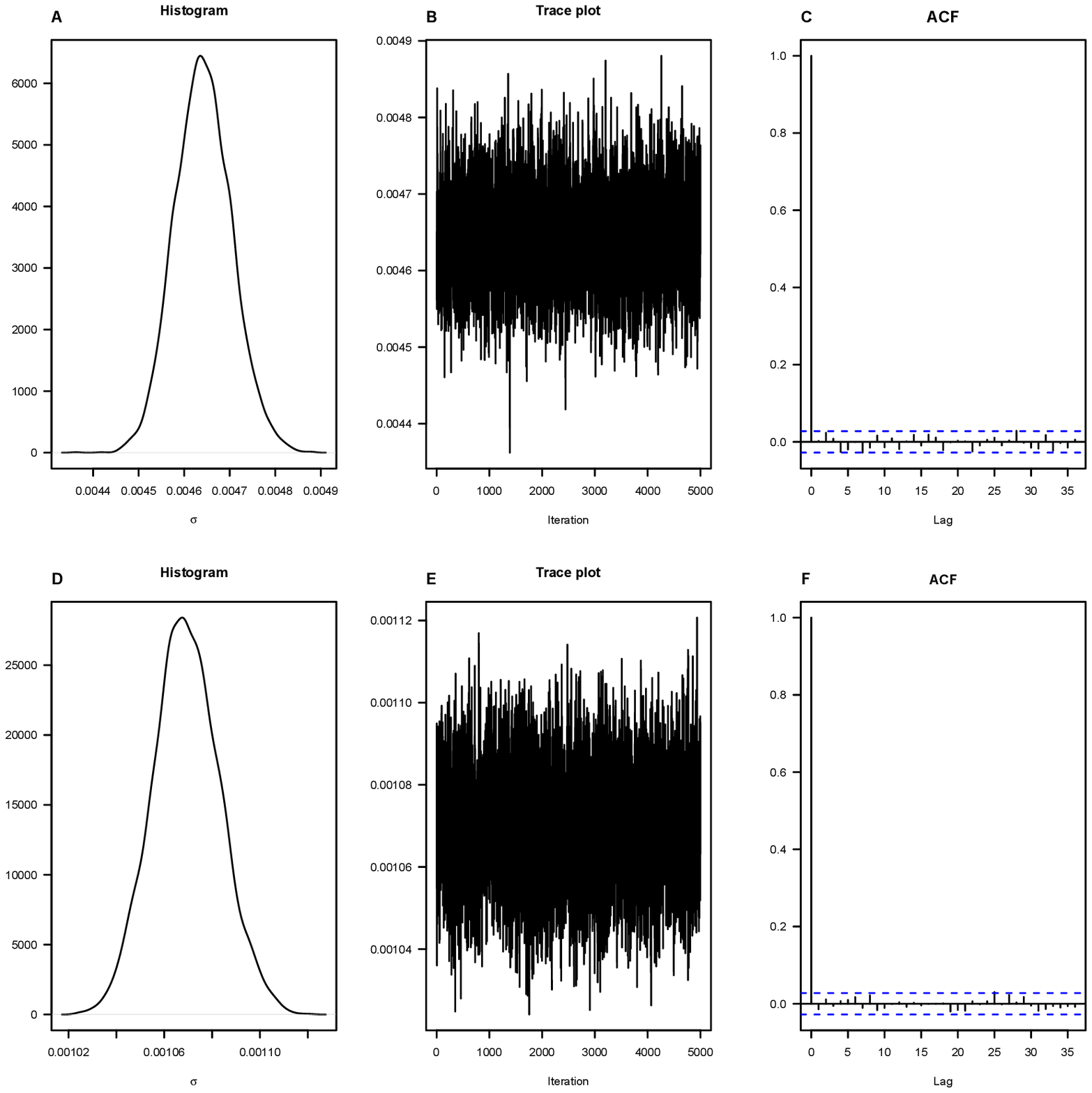}}
\caption{\label{fig:BohkRau_GBR_DNK_loglog_inSample}Marginal posterior density function (A, D), traceplot (B, E) and autocorrelation function (C, F)
for the parameter $\sigma$ in the British (top: A--C) and Danish (bottom: D--F) retrospective forecast with our model after $5,000$ iterations.} 
\end{figure}

As an example, Figure \ref{fig:BohkRau_GBR_DNK_loglog_inSample} depicts the marginal posterior density function,
the traceplot as well as the autocorrelation function exemplarily for the single model parameter $\sigma$,
the within age-group variance, of the British (upper panels: A--C) and Danish (lower panels: D--F) retrospective forecast.
The marginal density functions and the traceplots indicate that $\sigma$ converges to a stationary level,
i.\,e.~the parameter values oscillate around a constant value so that we can assume that the parameter space has been explored exhaustively.
The autocorrelation functions indicate that $\sigma$ is relatively good mixing because they decrease exponentially and stay at a low level for higher lags \citep{King2012}. \\

Table \ref{tab:RafteryLewis} lists the Raftery and Lewis's diagnostic;
i.\,e.~the number of iterations needed for each model parameter to obtain an accurrate estimate for the $0.025$ quantile with a probability of $0.95$.
For Great Britain as well as for Denmark, the estimated run-length \emph{N} fluctuates between $3,620$ and $3,741$ for all model parameters,
and the dependence factor~\emph{I} fluctuates between $0.966$ and $0.999$.
As neither the run-length \emph{N} nor the dependence factor \emph{I} have exceptionally high or low values for certain model parameters,
we conclude that $5,000$ iterations will be sufficient to estimate an accurate projection outcome with our proposed model.
As our model passes all these checks, we consider its results to be reliable.\\

\begin{table}
\caption{\label{tab:RafteryLewis}Raftery-Lewis diagnostic for single parameters of our retrospective mortality forecasts for women in Great Britain and Denmark.}
\centering
\fbox{%
\begin{tabular}{*{5}{c}}
 & & $\sigma$ & $\sigma_1$ & $\sigma_2$ \\
\hline
Great Britain & N &$3,741$ & $3,680$ & $3,620$  \\
 & I & $0.999$ & $0.982$ & $0.966$  \\
\hline
Denmark & N & $3,680$ & $3,741$ & $3,653$  \\
 & I & $0.982$ & $0.999$ & $0.975$  \\
\end{tabular}}
\end{table}

\section{Discussion}
\label{sec:discussion}

Advances in mortality forecasts will become even more important in the future than they are already today. 
The increase in life expectancy at birth in many highly developed countries is due to mortality reductions at all ages,
though highest gains in mortality reductions advance to successively older ages \citep{Christensen2009}.
Many mortality forecasts underestimate this ongoing progress in mortality reductions from younger to older ages and are, therefore, inaccurate.
In the past, this inaccuracy meant to underestimate mortality of people at working ages;
in the future it will underestimate mortality of people aged 65 and older. 
This will have an increasing impact on social welfare systems:
Underestimating mortality of people at working ages corresponds to having more people who pay premiums.
Underestimating mortality of people in retirement ages, however, will result in having more people who receive benefits for longer periods than previously anticipated.\\  

Common approaches, like the widely accepted Lee-Carter model and its numerous extensions,
typically extrapolate past mortality trends with an inflexible age schedule of mortality change and may induce, therefore, systematic forecast errors.
These forecast errors can become even larger when a country experiences an irregular mortality development.
We illustrate this effect by forecasting mortality for British and Danish women from 1991 to 2011 based on data from 1965 to 1990. 
While life expectancy at birth rose for British females at a regular pace,
it almost stagnated for Danish women in the 1980s and early 1990s.

Our comparison of the forecasted with the observed mortality data indicates that the original Lee-Carter model as well as its refinements proposed by Renshaw and Haberman \citeyearpar{Renshaw2003,Renshaw2006} could generate systematic forecast errors.
In fact, the retrospective mortality forecasts demonstrate that they systematically underestimate life expectancy at birth for British and Danish women:
In 2011, after only 21 forecast years, their absolute forecast error for Danish women is considerably larger than that for British women. \\

We address these issues in our probabilistic model with a novel combination of two modern concepts in mortality forecasting:
\begin{itemize}
\item We apply age-specific rates of mortality improvement to capture the dynamic age-shift of mortality change over time.
Rates of mortality improvement were employed in very recent studies by \cite{Mitchell2013} and by \cite{Haberman2012},
extending the predictor structure of the original Lee-Carter model. 
\item We optionally complement the mortality trend of a country of interest with that of at least one reference country to generate plausible forecasts even for countries with an irregular mortality development.
Recent works by \cite{Li2005} or \cite{Cairns2011} show that it can be worthwile to forecast mortality for at least two populations jointly,
especially when they exhibit similar characteristics concerning health and mortality.   
\end{itemize}

In contrast to the original model of Lee and Carter and its refinements proposed by Renshaw and Haberman \citeyearpar{Renshaw2003,Renshaw2006},
our model produces forecast errors that fluctuate around zero for life expectancy at birth of British and Danish women in the retrospective forecast.
In the case of Great Britain,
our model generates more accurate forecasts due to the rates of mortality improvement,
which allow a flexible age pattern of mortality change.  
Only the $P$-spline approach, proposed by \cite{curriemortality}, forecasts British mortality with forecast errors so small that they are similar to those of our model.
In the case of Denmark,
our model generates more accurate forecasts not only due to the application of the rates of mortality improvement,
but also due to the complement of the Danish mortality trend with that of Swedish women.
We decided to supplement the Danish with the Swedish mortality trend in the long run,
because we expected an acceleration in the sluggish trend of the base period for the forecast years.
Only the coherent Lee-Carter model, proposed by \cite{Li2005}, forecasts Danish mortality (also including the trend of Swedish women) with comparably small forecast errors as our model.
These findings suggest that our model can have a good forecasting performance for countries with various mortality conditions.  \\

In summary, we propose with our model a general framework,
which can be applied to derive mortality forecasts for any population---irrespective if a population experiences regular or irregular mortality developments.
This is because our model provides a wide range of distinctive methodological features that can be systematically applied---like,
for instance, the exchangeable core models (linear or log-log) or the option to include mortality trends of reference countries.
Although the results are not shown here, we gained further evidence from additional retrospective mortality forecasts for women and men that our model generates more accurate results than many other approaches not only for Great Britain and Denmark,
but also for Italy, Spain and West Germany.
These results are available for the user upon request to the authors. \\

The subjective choice of reference countries is certainly worth discussing.
Although purely data-driven methods are objective, they lack the ability to forecast trends that do not only extrapolate past developments.
To circumvent this problem we implemented the possibility to include subjective expert judgment in our model. 
If mortality trajectories are not expected to continue in a country of interest,
we suggest to supplement its mortality trend with those of at least one reference country.
Regarding the selection of appropriate reference countries,
we recommend to choose countries with morbidity and mortality conditions that are similar to those in the country of interest.

\section*{Acknowledgments}

The European Research Council has provided financial support under the European Community's Seventh Framework Programme (FP7/2007-2013) / ERC grant agreement no. 263744.
We would like to thank Ronald D. Lee, Webb Sprague and Carl Boe for their support to estimate the coherent Lee-Carter model. 
We would also like to thank the attendees of the Joint Eurostat/UNECE Work Session on Demographic Projections held in Rome, 29-31 October 2013, for their valuable comments.
 
\clearpage

\section*{Appendix}
\label{sec:appendix}

\begin{figure}[!ht]
\centering
\makebox{\includegraphics[width=0.8\textwidth]{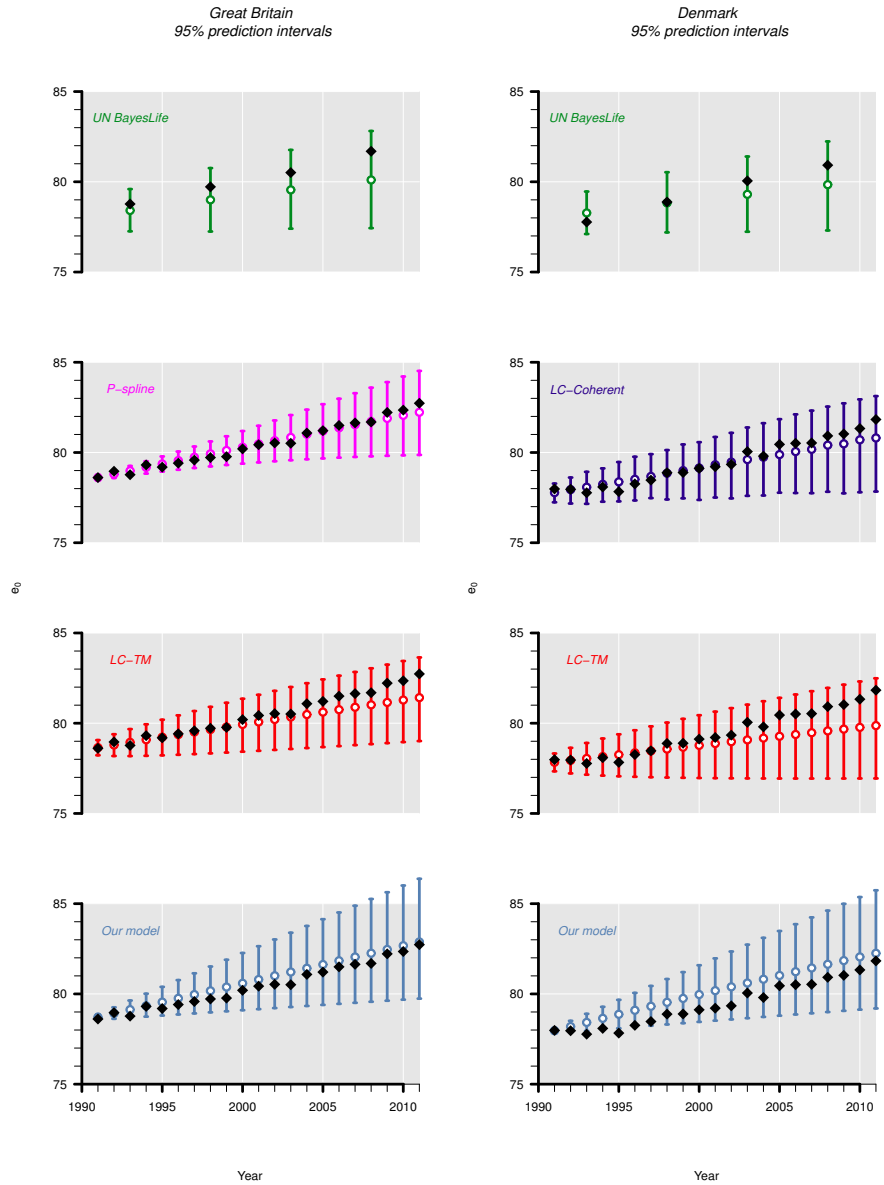}}
\caption{\label{fig:PredictionIntervals_DiffApproaches_DNKGBR_1965_1990_2011}Retrospective mortality forecasts of our model (blue), of the Lee-Carter model (red),
of the $P$-spline method (magenta), of the LC-coherent method (darkblue) and of the UN Bayesian approach (green) for British (left) and Danish (right) women.
Observed life expectancy at birth ($e_0$) is represented by black squares,
the median forecasts are represented by white circles and the width of the 95\% prediction intervals is represented by the length of the respective vertical lines.
In these retrospective forecasts, we take the data from 1965 to 1990 as basis to forecast mortality from 1991 to 2011.
Moreover, we complement the mortality trend for Danish women with that of Swedish women in our model as well as in the coherent Lee-Carter model.}
\end{figure}

\begin{figure}[!ht]
\centering
\makebox{\includegraphics[height=0.7\textheight]{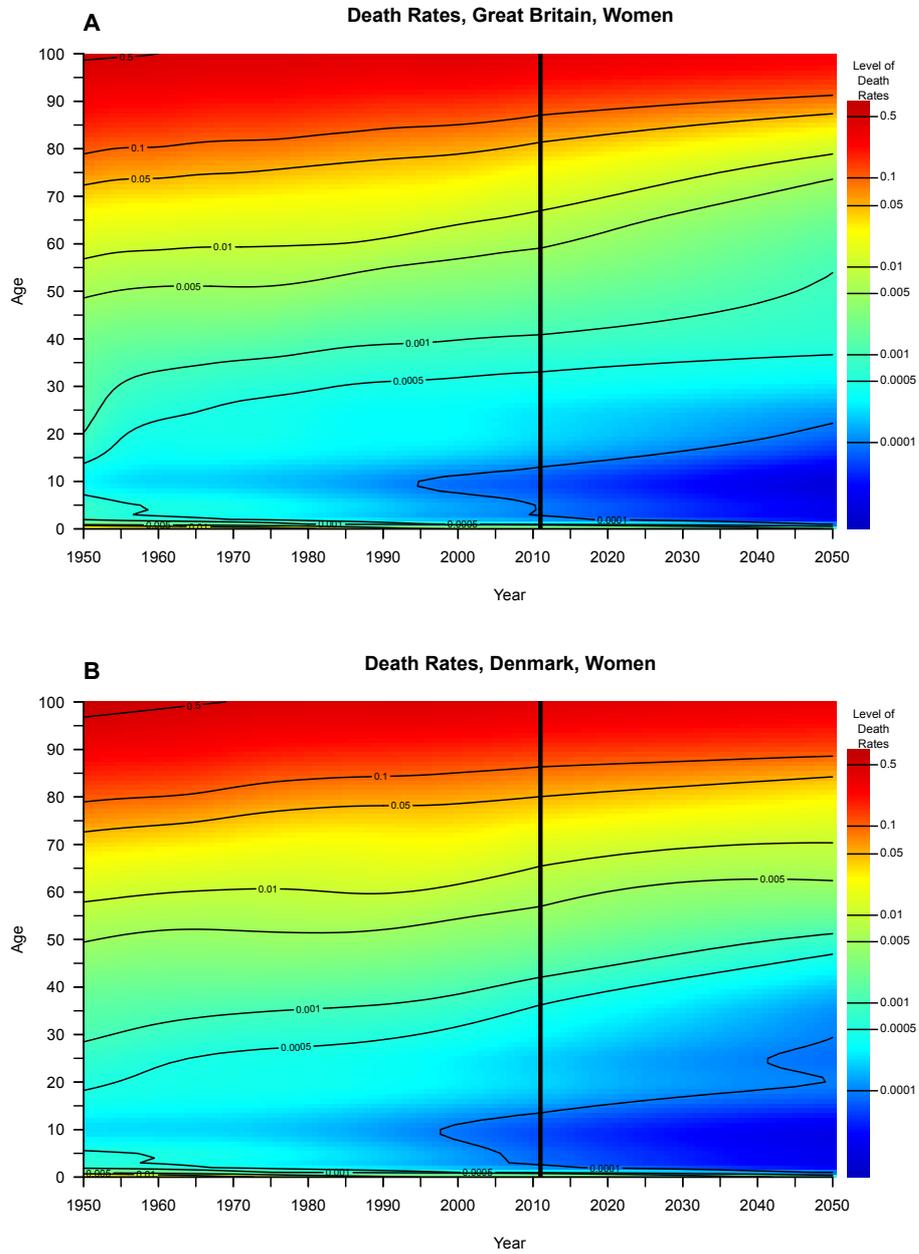}}
\caption{\label{fig:SurfacesLinearDeathRates1950.2050GBRandDNKwomen}
Death rates for British and Danish women from 1950 to 2050 of our proposed Bayesian linear model.
Observed death rates by single age from 1950 to 2011 and forecasted death rates using our proposed Bayesian linear model from 2012 to 2050
for women in Great Britain (A) and Denmark (B).
The transition from observed to forecasted death rates is illustrated by the vertical black line in both displays.} 
\end{figure}

\clearpage

\bibliographystyle{chicago}
\bibliography{BohkRau_bibliographyABBREV}

\end{document}